\documentclass{ws-procs9x6}

\newcommand{\refeq}[1]{(\ref{#1})}
\def\etal {{\it et al.}}

\begin{document}
\title{CONSTRAINING LORENTZ INVARIANCE VIOLATION WITH \textit{FERMI}}
\author{V.\ VASILEIOU for the \textit{Fermi} GBM and LAT Collaborations}
\address{NASA Goddard Space Flight Center \\ \& University of Maryland, Baltimore County\\
8800 Greenbelt Road,
Greenbelt, MD 20771, USA\\
E-mail: vlasios.vasileiou@nasa.gov}

\begin{abstract}
One of the cornerstones of special relativity is the postulate that all observers measure exactly the same photon speeds independently of the photon energies. However, a hypothesized structure of spacetime may alter this conclusion at ultra-small length scales. Even a tiny energy-dependent variation in the speed of light may be revealed, when accumulated over cosmological light-travel times, by high temporal-resolution measurements of sharp features in gamma-ray burst (GRB) lightcurves. We report the results of a study of the emission from GRB~090510 as detected by \textit{Fermi}'s LAT and GBM instruments that set unprecedented limits on the dependence of the speed of light on its energy. 
\end{abstract}

\bodymatter

\section{Introduction}
One of the predicted manifestations of Lorentz invariance violation (LIV) is a dependence of the speed of light in vacuo on its energy (see Refs.\ \refcite{liv1,liv5} and references therein). According to postulated LIV effects, two photons of energies $E_h>E_l$ emitted simultaneously from a distant astrophysical source at redshift $z$ will travel with different velocities and will arrive with a time delay $\Delta t$ equal to\cite{JacobPiran}:
\begin{equation}
\label{eqDT}
\Delta t=s_{n}\frac{(1+n)}{2H_{0}}\frac{(E_{h}^{n}-E_{l}^{n})}{(M_{QG,n}c^{2})^{n}}\int_{0}^{z}\frac{(1+z')^{n}}{\sqrt{\Omega_{m}(1+z')^{3}+\Omega_{\Lambda}}}dz',
\end{equation}
where $M_{QG,n}$ is the `quantum-gravity (QG) mass,' a parameter that sets the energy scale at which the QG effects that cause LIV start to become important. Its value is assumed to be near the Planck mass ($M_{Planck}\equiv \hbar c/ \lambda_{Planck} \sim 10^{19}$ GeV/$c^2$) and most likely smaller than it. The model-dependent parameter $n$ is assumed to be one or two, corresponding to linear ($\Delta t\propto \Delta E/M_{QG,1}$, with $\Delta E\equiv E_{h}-E_{l} \simeq E_{h}$) and quadratic ($\Delta t\propto \left(E_{h}/M_{QG,2}\right)^2$) LIV respectively. The model-dependent parameter $s_n$ is equal to plus or minus one, corresponding to a speed retardation or acceleration with an increasing photon energy respectively. In Ref.\ \refcite{nature} and using the above parametrization for $\Delta t$, we have placed constraints on LIV-induced dispersion in the form of lower limits on the quantum-gravity mass. In this proceeding, we are also constraining LIV using the parametrization of the Standard-Model Extension (SME)\cite{SME}.

\begin{figure}[ht!]
\label{fig:ET}
\centering
\includegraphics[width=0.7\columnwidth, height=12cm]{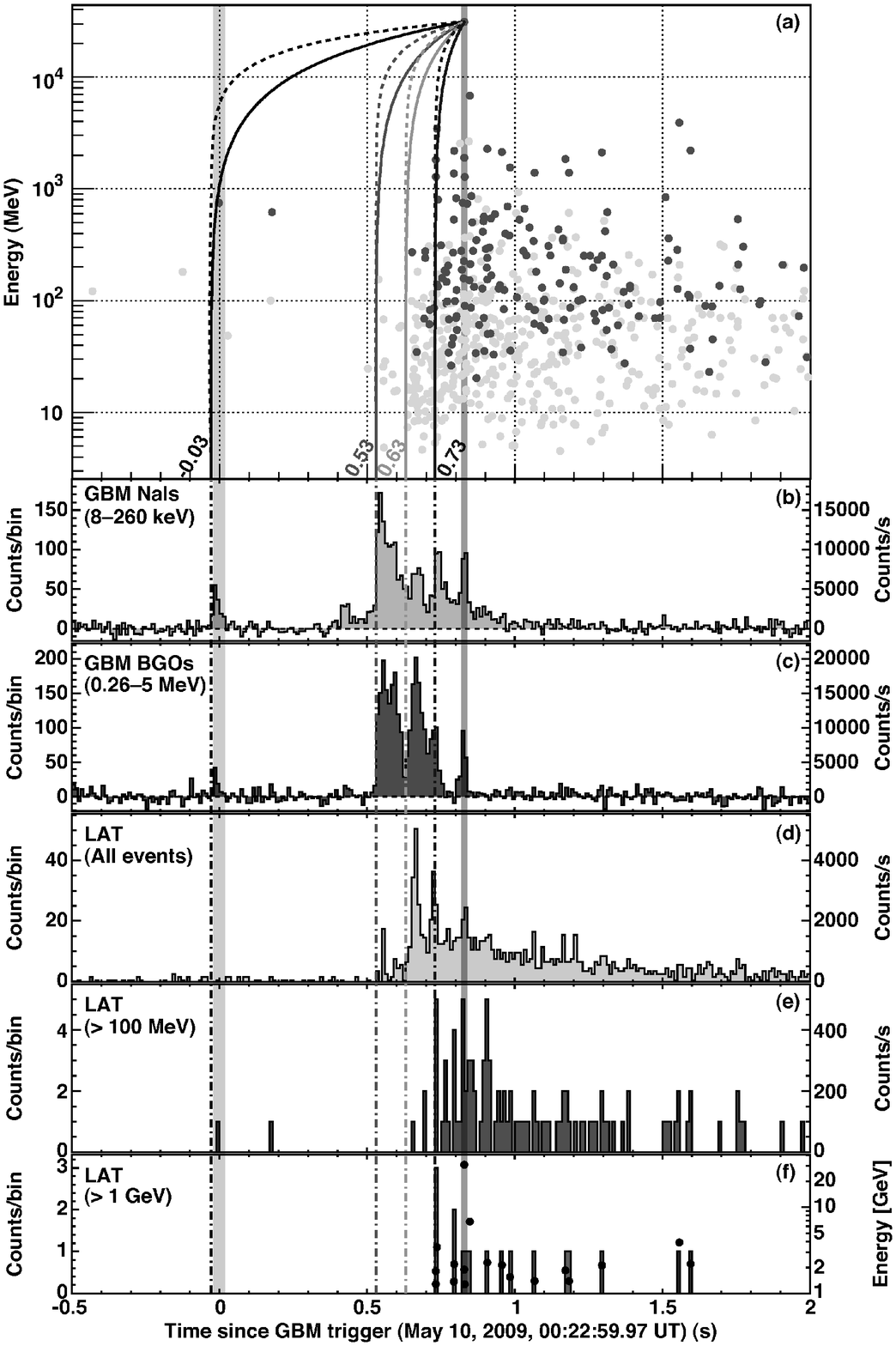}
\caption{GRB~090510 as observed by \textit{Fermi}.}
\end{figure}

Because of their short duration, rapid variability, and cosmological distances, GRBs are well-suited for constraining LIV. We used measurements on the bright and short GRB~090510 ($z=0.903\pm0.003$), which triggered both the LAT\cite{LAT} and GBM\cite{GBM} detectors on board \textit{Fermi} Gamma-Ray Space Telescope. The GBM and LAT lightcurves are shown in panels b--f of Fig.\ \ref{fig:ET}. The emission detected by the LAT extended to an energy of about 31 GeV (specifically $30.53^{+5.78}_{-2.57}$ GeV with 1$\sigma$ errors) (panel a of Fig.\ \ref{fig:ET}). The fact that this 31 GeV photon was detected shortly after the beginning of the burst ($\sim$0.8 s), and that the LAT-detected emission exhibited a series of very narrow spikes that extended to high (hundreds of MeV) energies, allowed us to set stringent limits on any LIV-induced dispersion effects as described below.

\section{Results}
To constrain any LIV-induced time delays we associated the 31 GeV photon with a lower-energy (LE) emission episode, during which we assumed that the 31 GeV photon was emitted. Such an association set an upper limit on any propagation time delay of the 31 GeV photon that was equal to the difference between the 31-GeV-photon's detection time ($t_{31 GeV}$=0.829 s) and the time of the LE-episode's start: ${\Delta}t_{max}=t_{31 GeV}-t_{start}$. This upper limit was then converted to a lower limit on $M_{QG,n}$ using Eq.\ \refeq{eqDT}. To be conservative in our calculations we used values for the 31 GeV photon's energy and the GRB's redshift reduced by 1$\sigma$. The most conservative and of a very-high-confidence assumption that can be made regarding the possible emission time of the 31 GeV photon is that it was not emitted before the beginning of the GRB (30 ms before the trigger). For such an assumption ${\Delta}t=0.859$ s, which corresponds to $M_{QG,1}\gtrsim1.19{\times}M_{Planck}$ and $M_{QG,2}\gtrsim2.99\times10^{10}\,GeV/c^2$. By associating the 31 GeV photon with other LE-emission episodes, as illustrated by the vertical lines in panels b--f of Fig.\ \ref{fig:ET}, we also produced more stringent yet with less confidence upper limits, reported in Table~\ref{table_old}. Our limits in the context of the SME and using the very high degree of confidence association with the E$>$100 MeV emission ($t_{start}=0.199$ s) are shown in Table~\ref{table_SME}. It should be noted that all limits produced by this method correspond to the case of $s_{n}=+1$ (light speed retardation with an increasing energy).
\begin{table}[ht]
\tbl{Our limits on $M_{QG,1}$ and $M_{QG,2}$.}
{\begin{tabular}{@{}clcc@{}}
\toprule
Limit on $\Delta t$ & Associated LE& Limit on $M_{QG,1}$ & Limit on $M_{QG,2}$\\
(ms) &emission episode&   ($M_{Planck}$) & ($10^{10}$ GeV/c$^{2}$)\\\colrule
$<$859&any $<$MeV emission&$>$1.19&$>$2.99\\
$<$299&main $<$MeV emission&$>$3.42&$>$5.06\\
$<$199&main $>$100 MeV emission&$>$5.12&$>$6.20\\
$<$99 &main $>$1 GeV emission&$>$10.0&$>$8.79\\
\multicolumn{2}{c}{DisCan: $|\Delta t/\Delta E|<$30 ms/GeV}&$>$1.22&--\\\botrule
\end{tabular}}
\label{table_old}
\end{table}

\begin{table}[ht]
\tbl{Our preliminary limits in the context of the SME framework.}
{\begin{tabular}{@{}ccc@{}}
\toprule
Model  & Coefficients & Upper Limits\\\colrule
Vacuum anisotropic  & $\sum_{jm}{}_0Y_{jm}\left(116^{\circ},334^{\circ}\right)c_{(I)jm}^{(6)}$ & $<3.9\times10^{-22}$ GeV$^{-2}$\\
&  $\sum_{jm}{}_0Y_{jm}\left(116^{\circ},334^{\circ}\right)c_{(I)jm}^{(8)}$ & $<2.1\times10^{-25}$ GeV$^{-4}$\\
Vacuum isotropic  & $c_{(I)00}^{(6)}$ & $<1.4\times10^{-21}$ GeV$^{-2}$\\
&  $c_{(I)00}^{(8)}$ & $<7.6\times10^{-25}$ GeV$^{-4}$\\\botrule
\end{tabular}}
\label{table_SME}
\end{table}

\begin{figure}[ht]
\centering
\includegraphics[width=0.65\columnwidth]{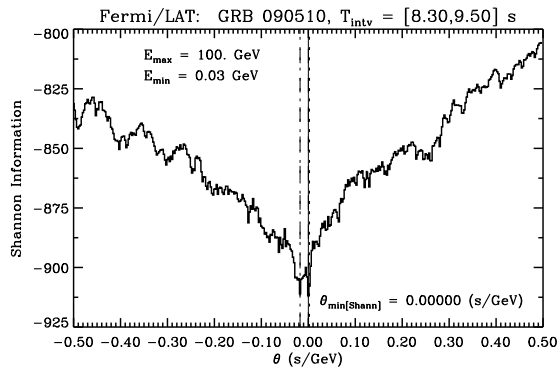}
\label{fig:DisCan1}
\caption{A measure of the LAT's lightcurve sharpness versus the trial spectral-lag value.} 
\end{figure}

We also used an alternative and independent method to constrain any linear-in-energy dispersion effects (DisCan method\cite{DisCan}). This method extracts dispersion information from all the LAT-detected photons ($\sim$30~MeV--$\sim$30 GeV) and is based on the fact that any QG-induced time delays would smear the spiky structure of the lightcurve. It applies trial spectral lags to the detection time of each detected photon to find the spectral lag that maximizes the sharpness of the lightcurve. The spectral lag that accomplishes that is equal and opposite in sign to the sum of any LIV-induced and intrinsic-to-the-GRB spectral lags. Figure \ref{fig:DisCan1} shows the value of a measure of the sharpness of the LAT lightcurve versus the trial spectral-lag value. The minimum of the curve, which denotes the most probable spectral lag, was at a value of $0^{+2}_{-18}$ ms/GeV. The errors here correspond to the trial spectral-lag values that were 100 times less probable than the best value of 0 ms/GeV and are shown with the two vertical dashed lines in the same figure. To estimate the inherent uncertainties due to our choice of time interval and energy range, and due to the limited statistics of the dataset, a bootstrap analysis was performed, in which the DisCan method was applied to multiple data sets produced by randomizing the association between the energies and times of the actually-detected events. The spectral lags of the randomized datasets, as measured by our method, lied within a value of $<$30 ms/GeV in 99\% of the cases. Therefore, our final combined result is a \textit{symmetric} upper limit on the energy dispersion equal to $|\Delta t/\Delta E|<$30 ms/GeV which corresponds to an upper limit of $M_{QG,1}>1.22 \times M_{Planck}$ for linear energy dispersion of either sign $s_n=\pm 1$ at the 99\% C.L. 

\section{Conclusion}
Using the observations of GRB~090510 by the GBM and LAT, we constrained tiny variations on the speed of light in vacuo that are linear or quadratic in its energy. We used two independent methods to obtain conservative and unprecedented upper limits on the magnitude of such speed variations. Our results ($M_{QG,1}\gtrsim$ few$\times M_{Planck}$) strongly disfavor any models predicting linear-in-energy variations in the speed of light. 

\section*{Acknowledgments}
The $Fermi$ LAT Collaboration acknowledges support from a number of agencies and institutes for both development and the operation of the LAT as well as scientific data analysis. These include NASA and DOE in the United States, CEA/Irfu and IN2P3/CNRS in France, ASI and INFN in Italy, MEXT, KEK, and JAXA in Japan, and the K.A.~Wallenberg Foundation, the Swedish Research Council and the National Space Board in Sweden. Additional support from INAF in Italy and CNES in France for science analysis during the operations phase is also gratefully acknowledged.
The Fermi GBM Collaboration acknowledges the support of NASA in the United States and DRL in Germany.

\end{document}